\documentclass[preprint,12pt]{elsarticle}

\usepackage{graphicx}
\usepackage{amssymb}
\usepackage{amsmath}
\usepackage{braket}

\usepackage[ruled,lined]{algorithm2e}

\newcommand{\ie}{\emph{i.e.}}
\newcommand{\eg}{\emph{e.g.}}

\newcommand{\rmd}{\mathrm{d}}
\newcommand{\rme}{\mathrm{e}}
\newcommand{\rmi}{\mathrm{i}}

\newcounter{bla}

\journal{Computer Physics Communications}

\begin{document}

\begin{frontmatter}



  \title{Program for quantum wave-packet dynamics with time-dependent
    potentials}


\author[umea]{C. M. Dion\corref{author}}
\author[umea]{A. Hashemloo}
\author[umea,jazan]{G. Rahali}

\cortext[author]{Corresponding author.\\\textit{E-mail address:}
  claude.dion@physics.umu.se} 
\address[umea]{Department of Physics, Ume{\aa} University, SE-901\,87 Ume{\aa}, Sweden}
\address[jazan]{Department of Physics, Jazan
  University, Jazan, Kingdom of Saudi Arabia.}

\begin{abstract}
  We present a program to simulate the dynamics of a wave packet
  interacting with a time-dependent potential.  The time-dependent
  Schr\"{o}dinger equation is solved on a one-, two-, or three-dimensional
  spatial grid using the split operator method. The program can be
  compiled for execution either on a single processor or on a
  distributed-memory parallel computer.
\end{abstract}


\end{frontmatter}



{\bf PROGRAM SUMMARY}

\begin{small}
\noindent
{\em Manuscript Title:} Program for quantum wave-packet dynamics with
time-dependent potentials                                        \\
{\em Authors:} C. M. Dion, A. Hashemloo, and G. Rahali \\
{\em Program Title:} wavepacket                                         \\
{\em Journal Reference:}                                      \\
{\em Catalogue identifier:}                                   \\
{\em Licensing provisions:} None                                   \\
{\em Programming language:} C (\textsc{iso} C99)                          \\
{\em Computer:}   Any computer with an \textsc{iso} C99 compiler (\eg,
gcc~\cite{1})\\
{\em Operating system:}  Any                                     \\
{\em RAM:} Strongly dependent on problem size.  See text for memory
estimates.                                               \\
{\em Number of processors used:} Any number from 1 to the number of
grid points along one dimension.                     \\
{\em Supplementary material:}                                 
{\em Classification:} 2.7 Wave Functions and Integrals  \\
{\em External routines/libraries:}   \textsc{fftw}~\cite{2}, \textsc{mpi} (optional)~\cite{3}  \\
{\em Subprograms used:} User-supplied potential function and routines
for specifying the initial state and optional user-defined observables.
\\
{\em Nature of problem:}\\
Solves the time-dependent Schr\"{o}dinger equation for a single particle
interacting with a time-dependent potential.
\\
{\em Solution method:}\\
The wave function is described by its value on a spatial grid and the
evolution operator is approximated using the split-operator method~\cite{4,5},
with the kinetic energy operator calculated using a Fast Fourier
Transform.
\\
{\em Restrictions:}\\
   \\
{\em Unusual features:}\\ 
Simulation can be in one, two, or three
dimensions.  Serial and parallel versions are compiled from the same
source files.
   \\
{\em Additional comments:}\\
   \\
{\em Running time:}\\
Strongly dependent on problem size.
   \\

\end{small}

\section{Introduction}
\label{sec:intro}

Quantum wave-packet dynamics, that is, the evolution of the spatial
distribution of a quantum particle, is an important part of the
simulation of many quantum systems.  It can be used for studying
problems as diverse as scattering, surface adsorption, and laser
control, just to name a few.

We propose here a general-purpose program to solve the spatial part of
the time-dependent Schr\"{o}dinger equation (\textsc{tdse}), aimed
particularly at a quantum particle interacting with a time-dependent
potential.  Our interest mainly concerns such applications as laser
control of quantum
systems~\cite{Letokhov_book_2007,Shapiro_book_2012}, but the program
can be used with any user-supplied potential function.

The program is based on the split-operator
method~\cite{split:fleck76,fft:feit82,fft:feit83,split:bandrauk91},
which has successfully been used to solve the time-dependent
Schr\"{o}dinger equation in many different settings, from the calculation
of vibrational bound states (see, \eg, \cite{fft:feit83}) and the
simulation of high-power laser-matter interactions (see, \eg,
\cite{Salamin_PR_2006}), to the laser control of chemical reactions
(see, \eg, \cite{iso:dion96}).  The method can also be applied to
Schr\"{o}dinger-like equations, such as the
Gross-Pitaevskii~\cite{gp:javanainen06} and
Dirac~\cite{split:mocken08} equations.

\section{Numerical approach}
\label{sec:num}

\subsection{Split-operator method}

In this section, we present a detailed description of the
split-operator method to solve the time-dependent Schr\"{o}dinger
equation.  While everything presented here can be found in the
original works developing the
method~\cite{split:fleck76,fft:feit82,fft:feit83,split:bandrauk91}, we
think it useful to review all the elements necessary to understand the
inner workings of the program.

We consider the time-dependent Schr\"{o}dinger equation,
\begin{equation}
  \rmi\hbar \frac{\partial}{\partial t} \psi(t) = \hat{H}
  \psi(t),
  \label{eq:tdse}
\end{equation}
with $\hat{H}$ the Hamiltonian for the motion of a particle
interacting with an external time-dependent potential $V(t)$, \ie,
\begin{equation}
  \hat{H} = \hat{K} + \hat{V} = \frac{\hat{P}^2}{2 m} + V(t),
  \label{eq:hamiltonian}
\end{equation}
where $\hat{K}$ and $\hat{V}$ are the kinetic and potential energy
operators, respectively, $\hat{P}$ is the momentum operator, and $m$
the mass of the particle.  (The same Hamiltonian is obtained for a
vibrating diatomic molecule, where the spatial coordinate is replaced
by the internuclear distance, and the potential $V(t)$ is the sum of
the internal potential energy and an external, time-dependent
potential, as will be shown in Sec.~\ref{sec:vib}.)

The formal solution to eq.~(\ref{eq:tdse}) is given by the time
evolution operator $\hat{U}$, itself a solution of the time-dependent
Schr\"{o}dinger equation~\cite{cohen-tannoudji92},
\begin{equation}
  \rmi\hbar \frac{\partial}{\partial t} \hat{U} = \hat{H} \hat{U},
\end{equation}
such that, given an initial wave function at time $t_0$, $\psi(t_0)$,
the solution at any time $t$ is obtained from 
\begin{equation}
  \psi(t) = \hat{U} (t,t_0) \psi(t_0).
\end{equation}
As the Hamiltonian is time dependent, we have that~\cite{oqs:breuer02}
\begin{align}
  \hat{U} (t,t_0) &= \hat{T} \exp \left[ -\frac{\rmi}{\hbar}
    \int_{t_0}^{t} \hat{H}(t') \rmd t' \right] \nonumber \\
  &= \hat{T} \exp \left\{ -\frac{\rmi}{\hbar} \int_{t_0}^{t} \left[
      \hat{K} + \hat{V}(t') \right] \rmd t' \right\}.
\label{eq:Ut0t}
\end{align}
In eq.~(\ref{eq:Ut0t}), the time-ordering operator $\hat{T}$ ensures
that the Hamiltonian is applied to the wave function in order of
increasing time, as in general the Hamiltonian does not commute with
itself at a different time, \ie, $[\hat{H}(t), \hat{H}(t')] \neq 0$ iff $t \neq
t'$~\cite{cohen-tannoudji92,method:pechukas66}.  By considering a
small time increment $\Delta t$, we can do without the time-ordering
operator by considering the approximate short-time evolution
operator~\cite{method:pechukas66},
\begin{equation}
  \hat{U}(t + \Delta t,t) = \exp \left\{ -\frac{\rmi}{\hbar}
    \int_{t}^{t+\Delta t} \left[ 
      \hat{K} + \hat{V}(t') \right] \rmd t' \right\}.
  \label{eq:short-time}
\end{equation}

We are concerned here with time-dependent potentials that also have a
spatial dependence, $\hat{V} \equiv V(\mathbf{x},t)$, such as those
produced by ion traps or focused laser pulses, such that $\hat{V}
\equiv V(\mathbf{x},t)$, in which case $\hat{K}$ and $\hat{V}$ do not
commute.  For two non-commuting operators $\hat{A}$ and $\hat{B}$,
$\rme^{\hat{A} + \hat{B}} \neq \rme^{\hat{A}} \rme^{\hat{B}}$, but the
split-operator method~\cite{fft:feit82,fft:feit83} allows the
approximation of the evolution operator with minimal error,
\begin{align}
  \hat{U}(t + \Delta t,t) &= \exp \left[ -\frac{\rmi \Delta t}{2 \hbar}
    \hat{K} \right] \exp \left[ -\frac{\rmi}{\hbar} \int_{t}^{t+\Delta
      t} \hat{V}(t') \rmd t' \right] \nonumber \\
  &\quad \times \exp \left[ -\frac{\rmi \Delta t}{2 \hbar} \hat{K}
  \right] + O(\Delta t^3). 
\end{align}
Using the midpoint formula~\cite{num:conte72} for the integral of the potential,
\begin{equation}
  \int_{t}^{t + \Delta t} f(t') \rmd t' = f(t + \Delta t/2) \Delta t +
  O(\Delta t^3),
\end{equation}
we get 
\begin{equation}
  \hat{U}(t + \Delta t,t) \approx \exp \left[ -\frac{\rmi \Delta t}{2
      \hbar} \hat{K} \right] \exp \left[ -\frac{\rmi \Delta t}{\hbar}
    V(t+ \frac{\Delta t}{2}) \right] \exp \left[ -\frac{\rmi \Delta
      t}{2 \hbar} \hat{K} \right],
  \label{eq:splitop} 
\end{equation}
where the global error is $O(\Delta t^3)$.  The choice of the order of
the operators $\hat{K}$ and $\hat{V}$ in the above equations is
arbitrary, but the choice we make here allows for a faster execution
in the majority of cases, \ie, when the intermediate value of the wave
function is not needed at all time steps.  We can then link together
$n$ consecutive time steps into
\begin{align}
  \hat{U}(t + n \Delta t,t) &= \hat{U}(t + n \Delta t, t + [n-1]
  \Delta t) \hat{U}(t + [n-1] \Delta t, t + [n-2] \Delta t) \nonumber
  \\
  &\quad \times \cdots \times
  \hat{U}(t + \Delta t, t)   \nonumber \\
  &= \exp \left[ -\frac{\rmi \Delta t}{2
      \hbar} \hat{K} \right] \exp \left[ -\frac{\rmi \Delta t}{\hbar}
    \hat{V}( t + \frac{2n-1}{2}\Delta t) \right] \nonumber \\
  &\quad \times \left\{ \prod_{j=n-1}^{1} \exp \left[ -\frac{\rmi
        \Delta t}{\hbar} 
      \hat{K} \right] 
    \exp \left[ -\frac{\rmi \Delta t}{\hbar}
    \hat{V}( t + \frac{2j-1}{2}\Delta t) \right] \right\} \nonumber \\
 &\quad \times \exp \left[ -\frac{\rmi \Delta t}{2
      \hbar} \hat{K} \right],
\end{align}
where two sequential operations of $\hat{K}$ are combined into one.
The same is not possible with $\hat{V}$ due to its time dependence.

We choose to discretize the problem on a finite spatial grid, \ie,
$\textbf{x} = (x,y,z)$ is restricted to the values
\begin{align}
  x_i &= x_\mathrm{min} + i \Delta x, & i = 0, \ldots, n_x-1 , \nonumber \\
  y_j &= y_\mathrm{min} + j \Delta y, & j = 0, \ldots, n_y-1 , \nonumber \\
  z_k &= z_\mathrm{min} + k \Delta z, & k = 0, \ldots, n_z-1 ,
  \label{eq:grid}
\end{align}
where the number of grid points $(n_x, n_y, n_z)$ are (integer)
parameters, as is the size of the grid, with bounds $x \in
[x_\mathrm{min}, x_\mathrm{max}]$ and where
\begin{equation}
  \Delta x  = \frac{x_\mathrm{max} - x_\mathrm{min}}{n_x - 1},
  \label{eq:Deltax}
\end{equation} 
with equivalent expressions in $y$ and $z$.

The problem now becomes that of calculating the exponential of
matrices $\mathsf{K}$ and $\mathsf{V}$, which is only trivial for a
diagonal matrix~\cite{moler_SIAMRev_2003}.  In the original
implementation of the split-operator
method~\cite{fft:feit82,fft:feit83}, this is remedied by considering
that while the matrix for $\hat{V}$ is diagonal for a spatial
representation of the wave function, $\hat{K}$ is diagonal in momentum
space.  By using a Fourier transform (here represented by the operator
$\mathcal{F}$) and its inverse ($\mathcal{F}^{-1}$), we can write
\begin{equation}
  \exp \left[ -\frac{\rmi \Delta t}{2
      \hbar} \hat{K}(\mathbf{x}) \right] \psi(\mathbf{x}) =
  \mathcal{F}^{-1} \exp \left[ -\frac{\rmi \Delta t}{2
      \hbar} \hat{K}(\mathbf{p}) \right] \mathcal{F} \psi(\mathbf{x}),
  \label{eq:ft}
\end{equation}
where, considering that $\hat{K} = \hat{P}^2 / 2m$,
\begin{align}
  \hat{K}(\mathbf{p}) &= \frac{\mathbf{p}^2}{2 m} , \\
  \hat{K}(\mathbf{x}) &= -\frac{\hbar^2}{2m} \nabla^2,
\end{align}
since the operators transform as $-\rmi \hbar \nabla \Leftrightarrow
\mathbf{p}$ when going from position to momentum
space~\cite{cohen-tannoudji92}. Equation~(\ref{eq:ft}) is efficiently
implemented numerically using a Fast Fourier Transform
(\textsc{fft})~\cite{pressC92}. After the forward transform, the
momentum grid, obtained from the wave vector $\mathbf{k} =
\mathbf{p}/\hbar$, is discretized according to~\cite{pressC92}
\begin{align}
  p_{x,i} &= 2\pi \hbar \frac{i}{n_x \Delta x}, & i = -\frac{n_x}{2},
  \ldots, \frac{n_x}{2}, \nonumber \\
  p_{y,j} &= 2\pi \hbar \frac{j}{n_y \Delta y}, & j = -\frac{n_y}{2},
  \ldots, \frac{n_y}{2}, \nonumber \\
  p_{z,k} &= 2\pi \hbar \frac{k}{n_z \Delta z}, & k = -\frac{n_z}{2},
  \ldots, \frac{n_z}{2}.
\end{align}
Care must be taken to associate the appropriate momentum value to each
element of the Fourier-transformed wave function, considering the
order of the output from \textsc{fft} routines~\cite{pressC92}.
Algorithm~\ref{algo:split} summarizes the split-operator method as
presented here.
\begin{algorithm}
  \caption{\label{algo:split}Main algorithm for the split-operator
    method.}
  \DontPrintSemicolon
  Initialize $\psi(t=0)$\;
  \For{$j \leftarrow 1$ \KwTo $n_{\mathrm{t}} / n_{\mathrm{print}} $}
  { 
    $\tilde{\psi}(\mathbf{p}) \leftarrow \mathcal{F}
    \psi(\mathbf{x})$\;
    Multiply $\tilde{\psi}(\mathbf{p})$ by $\exp \left[ -\frac{\rmi
        \Delta t}{2 \hbar} \frac{\mathbf{p}^2}{2m} \right]$\;
    $\psi(\mathbf{x}) \leftarrow \mathcal{F}^{-1}
    \tilde{\psi}(\mathbf{p})$\;
    \For{$i \leftarrow 1$ \KwTo $n_{\mathrm{print}}-1$}
    {
      Multiply $\psi(\mathbf{x})$ by $\exp \left[ -\frac{\rmi
        \Delta t}{\hbar} V(\mathbf{x},t) \right]$\;
      $\tilde{\psi}(\mathbf{p}) \leftarrow \mathcal{F}
      \psi(\mathbf{x})$\;
      Multiply $\tilde{\psi}(\mathbf{p})$ by $\exp \left[ -\frac{\rmi
          \Delta t}{\hbar} \frac{\mathbf{p}^2}{2m} \right]$ \;
      $\psi(\mathbf{x}) \leftarrow \mathcal{F}^{-1} \tilde{\psi}(\mathbf{p})$\;
    }
    Multiply $\psi(\mathbf{x})$ by $\exp \left[ -\frac{\rmi
        \Delta t}{\hbar} V(\mathbf{x},t) \right]$\;
    $\tilde{\psi}(\mathbf{p}) \leftarrow \mathcal{F}
    \psi(\mathbf{x})$\;
    Multiply $\tilde{\psi}(\mathbf{p})$ by $\exp \left[ -\frac{\rmi
        \Delta t}{2 \hbar} \frac{\mathbf{p}^2}{2m}  \right]$\;
    $\psi(\mathbf{x}) \leftarrow \mathcal{F}^{-1}
    \tilde{\psi}(\mathbf{p})$\; 
    Calculate observables $\braket{\hat{A}} \equiv
    \braket{\psi(\mathbf{x}) | \hat{A} | \psi(\mathbf{x})}$\;
  }
\end{algorithm}

\subsection{Parallel implementation}

\label{sec:parallel}

We consider now the implementation of the algorithm described above on
a multi-processor architecture with distributed memory.  The
``natural'' approach to parallelizing the problem is to divide the
spatial grid, and therefore the wave function, among the processors.
Each processor can work on its local slice of the wave function,
except for the Fourier transform, which requires information across
slices. This functionality is pre-built into the parallel
implementation of the \textsc{fft} package
\textsc{fftw}~\cite{FFTW05}, of which we take advantage.  The
communications themselves are implemented using the Message Passing
Interface (\textsc{mpi}) library~\cite{mpi,mpi:gropp99}.

For a 3D (or 2D) problem, the wave function is split along the $x$
direction, with each processor having a subset of the grid in $x$, but
with the full extent in $y$ and $z$.  To minimize the amount of
communication after the forward \textsc{fft}, we use the intermediate
transposed function, where the split is now along the $y$ dimension.
The original arrangement is recovered after the backward function, so
this is transparent to the user of our program.  In addition,
\textsc{fftw} offers the possibility of performing a 1D transform in
parallel, which we also implement here.

The only constrain this imposes on the user is that a 1D problem may
only be defined along $x$, and a 2D problem in the $xy$-plane (in
order to simplify the concurrent implementation of serial and parallel
versions, this constraint also applies to the serial version).  In
addition to the total number of grid points along $x$, $n_x$, each
processor has access to $n_{x, \mathrm{local}}$, the number of grid
points in $x$ for this processor, along with $n_{x, 0}$, the
corresponding initial index.  In other words, each processor has a
grid in $x$ defined by
\begin{align}
  x_i &= x_\mathrm{min} + \left( i + n_{x, 0} \right) \Delta x, & i =
  0, \ldots, n_{x, \mathrm{local}},
\end{align}
with the grids in $y$ and $z$ still defined by eq.~(\ref{eq:grid}).

\section{User guide}
\label{sec:guide}

\subsection{Summary of the steps for compilation and execution}

Having defined the physical problem to be simulated, namely by setting
up the potential $V(\mathbf{x},t)$ and initial wave function
$\psi(\mathbf{x};t=0)$, the following routines must be coded (see
section~\ref{sec:user-def} for details):
\begin{itemize}
\item \texttt{initialize\_potential}
\item \texttt{potential}
\item \texttt{initialize\_wf}
\item \texttt{initialize\_user\_observe} (can be empty)
\item \texttt{user\_observe} (can be empty)
\end{itemize}
The files containing these functions must include the header file
\texttt{wavepacket.h}. The program can then be compiled according to
the instructions in section~\ref{sec:compiling}.

A parameter file must then be created, see
section~\ref{sec:parameters}.  The program can then be executed using
a command similar to
\begin{verbatim}
wavepacket parameters.in
\end{verbatim}

\subsection{User-defined functions}
\label{sec:user-def}

The physical problem that is actually simulated by the program depends
on two principal elements, the time-dependent potential $V(\mathbf{x},
t)$ and the initial wave function $\psi( \mathbf{x}; t=0)$.  In
addition, the user may be interested in observables that are not
calculated by the main program (the list a which is given in
Sec.~\ref{sec:parameters}).  The user \emph{must} supply functions
which define those elements, which are linked to at compile time.  How
these functions are declared and what they are expected to perform is
described in what follows, along with the data structure that is
passed to those functions.

\subsubsection{Data structure \texttt{parameters}}
\label{sec:struct_params}

The data structure \texttt{parameters} is defined in the header file
\texttt{wavepacket.h}, which must be included at the top of the users
own C files to be linked to the program.  A variable of type
\texttt{parameters} is passed to the user's functions, and contains
all parameters the main program is aware of and that are
useful/necessary for the execution of the tasks of the user-supplied
routines.  The structure reads
\begin{verbatim}
typedef struct
{
  /* Parameters and grid */
  int size, rank;
  size_t nx, ny, nz, n, nx_local, nx0, n_local;
  double x_min, y_min, z_min, x_max, y_max, z_max, dx, dy, dz;
  double *x, *y, *z, *x2, *y2, *z2;
  double mass, dt, hbar;
} parameters;
\end{verbatim}
where the different variables are:
\begin{itemize}
\item \texttt{size}: Number of processors on which the program is
  running.

\item \texttt{rank}: Rank of the local processor, with a value in the
  range $[0, \texttt{size}-1]$.  In the serial version, the value is
  therefore \texttt{rank = 0}.  (\emph{Note:} All input and output
  to/from disk is performed by the processor of \texttt{rank} 0.)

\item \texttt{nx, ny, nz}: Number of grid points along $x$, $y$, and
  $z$, respectively.  In the parallel version, this refers to the full
  grid, which is then split among the processors.  For a one or
  two-dimensional problem, \texttt{ny} and/or \texttt{nz} should be
  set to 1.  ($x$ is always the principal axis in the program.)  For
  best performance, these should be set to a product of powers of
  small prime integers, \eg,
  $$
  \texttt{nx} = 2^i 3^j 5^k 7^l.
  $$
  See the documentation of \textsc{fftw} for more details~\cite{fftw}.

\item $\texttt{n} = \texttt{nx} \times \texttt{ny} \times
  \texttt{nz}$.

\item \texttt{nx\_local}: Number of grid points in $x$ on the local
  processor, see Sec.~\ref{sec:parallel}.  In the serial version,
  \texttt{nx\_local = nx}.

\item \texttt{nx0}: Index of the first local grid point in $x$, see
  Sec.~\ref{sec:parallel}.  In the serial version, $\texttt{nx0} = 0$.

\item \texttt{x\_min, y\_min, z\_min, x\_max, y\_max, z\_max}: Values
  of the first and last grid points along $x$, $y$, and $z$.

\item \texttt{dx, dy, dz}: Grid spacings $\Delta x$, $\Delta y$, and
  $\Delta z$, respectively, see eq.~(\ref{eq:Deltax}).

\item \texttt{x, y, z}: Arrays of size \texttt{nx\_local},
  \texttt{ny}, and \texttt{nz}, respectively, containing the value of
  the corresponding coordinate at the grid point.

\item \texttt{x2, y2, z2}: Arrays of size \texttt{nx\_local},
  \texttt{ny}, and \texttt{nz}, respectively, containing the square of
  the value of the corresponding coordinate at the grid point.

\item \texttt{mass}: Mass of the particle.

\item \texttt{dt}: Time step $\Delta t$ of the time evolution, see
  eq.~(\ref{eq:short-time}).

\item \texttt{hbar}: Value of $\hbar$, Planck's constant over $2\pi$,
  in the proper units.  (See Sec.~\ref{sec:parameters}.)

\end{itemize}

\subsubsection{Initializing the potential}
\label{sec:init_pot}

In the initialization phase of the program, before the time evolution,
the function
\begin{verbatim}
void 
initialize_potential (const parameters params, const int argv, 
                      char ** const argc);
\end{verbatim}
is called, with the constant variable \texttt{params} containing all
the values specified in Sec.~\ref{sec:struct_params}.  \texttt{argv}
and \texttt{argc} are the variables relating to the command line
arguments, as passed to the main program:
\begin{verbatim}
int
main (int argv, char **argc);
\end{verbatim}

This function should perform all necessary pre-calculations and
operations, including reading from a file additional parameters, for
the potential function.  The objective is to reduce as most as
possible the time necessary for a call to the \texttt{potential}
function.

\subsubsection{Potential function}

The function
\begin{verbatim}
double 
potential (const parameters params, const double t, 
           double * const pot);
\end{verbatim}
should return the value of the potential $V(\mathrm{x},t)$, for all
(local) grid points at time \texttt{t}, in the array \texttt{pot}, of
dimension \texttt{pot[nx\_local][ny][nz]}.

\subsubsection{Initial wave function}

The initial wave function $\psi(\mathbf{x},t=0)$ is set by the
function
\begin{verbatim}
void 
initialize_wf (const parameters params, const int argv, 
               char ** const argc, double complex *psi);
\end{verbatim}
where \texttt{psi} is a 3D array of dimension
\texttt{psi[nx\_local][ny][nz]}.  If the wave function is to be read
from a file, users can make use of the functions
\texttt{read\_wf\_text} and \texttt{read\_wf\_bin}, described in
Sec.~\ref{sec:useful}.

\subsubsection{User-defined observables}

\label{sec:user}

In addition to the observables that are built in, which are described
in Sec.~\ref{sec:parameters}, users may define additional observables,
such as the projection of the wave function on eigenstates.

The function
\begin{verbatim}
void 
initialize_user_observe (const parameters params, const int argc, 
                         char ** const argv);
\end{verbatim}
is called once at the beginning of the execution.  It should perform
all operations needed before any call to \texttt{user\_observe}.  The
arguments passed to the function are the same as those of
\texttt{initialize\_potential}, see Sec.~\ref{sec:init_pot}.

During the time evolution, every \texttt{nprint} time step, the
function
\begin{verbatim}
void 
user_observe (const parameters params, const double t, 
              const double complex * const psi);
\end{verbatim}
is called, with the current time \texttt{t} and wave function
\texttt{psi}.

The printing out of the results, as well as the eventual opening of a
file, is to be performed within these user-supplied functions.  In a
parallel implementation, only the processor of \texttt{rank} 0 should
be responsible for these tasks, and proper communication must be set
up to ensure the full result is available to this processor.

Note that these functions \emph{must} be present in the source file
that will be linked with the main program, even if additional
observables are not desired.  In this case, the function body can be
left blank.

\subsubsection{Useful functions}
\label{sec:useful}

A series of functions declared in the header file
\texttt{wavepacket.h} and that are part of the main program are also
available for use within the user-defined functions described above.

\begin{itemize}

\item
\begin{verbatim}
double 
norm (const parameters params, 
      const double complex * const psi);
\end{verbatim}
calculates $\sqrt{\braket{\mathtt{psi} | \mathtt{psi}}}$.

\item
\begin{verbatim}
double complex 
integrate3D (const parameters params, 
             const double complex * const f1,
             const double complex * const f2);
\end{verbatim}
given $f_1 \equiv \mathtt{f1}$ and $f_2 \equiv \mathtt{f2}$,
calculates 
$$
\braket{f_1 | f_2} = 
\int_{z_{\mathrm{min}}}^{z_{\mathrm{max}}}
\int_{y_{\mathrm{min}}}^{y_{\mathrm{max}}}
\int_{x_{\mathrm{min}}}^{x_{\mathrm{max}}} f_1^* f_2 \, \rmd x \, \rmd
y \,
\rmd z.
$$
(Correct results are also obtained for 1D and 2D systems.)

\item
\begin{verbatim}
double 
expectation1D (const parameters params, const int dir, 
               const double * const f, 
               const double complex * const psi);
\end{verbatim}
given $f(\xi) \equiv \mathtt{f}$ and $\psi \equiv \mathtt{psi}$,
calculates
$$
\braket{\psi | f({\xi}) | \psi} = 
\int_{z_{\mathrm{min}}}^{z_{\mathrm{max}}}
\int_{y_{\mathrm{min}}}^{y_{\mathrm{max}}}
\int_{x_{\mathrm{min}}}^{x_{\mathrm{max}}} \psi^* f(\xi) \psi \, \rmd x \, \rmd
y \,
\rmd z,
$$
where $\xi = x, y, z$ for $\texttt{dir} = 1, 2, 3$, respectively.

\item
\begin{verbatim}
void 
read_wf_bin (const parameters params, 
             const char * const wf_bin,
             double complex * const psi);
\end{verbatim}
opens the file with filename \texttt{wf\_bin} and reads the wave
function into \texttt{psi}.  The file must be in a binary format, as
written when the keyword \texttt{wf\_output\_binary} is present in the
parameter file, see Sec.~\ref{sec:parameters}.  In the parallel
version, the file is read by the processor of rank 0, and each
processor is assigned its local part of the wave function of size
\texttt{psi[nx\_local][ny][nz]}.

\item
\begin{verbatim}
void
distribute_wf (const parameters params, 
               double complex * const psi_in,
               double complex * const psi_out);
\end{verbatim}
given the wave function \texttt{psi\_in[nx][ny][nz]} located on the
processor of rank 0, returns in \texttt{psi\_out[nx\_local][ny][nz]}
the local part of the wave function on each processor.  Intended only
to be used in the parallel version, the function will simply copy
\texttt{psi\_in} into \texttt{psi\_out} in the serial version.

\item
\begin{verbatim}
void
abort ()
\end{verbatim}
terminates the program.  This is the preferred method for exiting the
program (\eg, in case of error) in user-supplied routines, especially
in the parallel version.

\end{itemize}

\subsection{Compiling the program}
\label{sec:compiling}

A sample makefile is supplied with the program, which should be
straightforward to adapt to one's needs.  Without a makefile, a
typical command-line compilation would look something like
\begin{verbatim}
gcc -O3 -std=c99 -o wavepacket wavepacket.c user_defined.c \
    -lfftw3 -lm 
\end{verbatim}
where the file \texttt{user\_defined.c} contains all the routines
specified in section~\ref{sec:user-def}.

By default, the compiling will produce the serial version of the
program.  To compile the \textsc{mpi} parallel version requires
defining the macro \texttt{MPI}, \ie, by adding \texttt{-DMPI} as an
argument to the compiler (through \texttt{CFLAGS} in the makefile).
In addition, \textsc{mpi} libraries must be linked to, including
\texttt{-lfftw3\_mpi}.

\subsection{Parameter file}
\label{sec:parameters}

When executing the program, it will expect the first command-line
argument to consist of the name of the parameter file.  This file is
expected to contain a series of statements of the type '\emph{key} =
\emph{value}', each on a separate line.  The order of these statements
is not important, and blank lines are ignored, but white space must
separate \emph{key} and \emph{value} from the equal sign.  Note that
the program does not check for duplicate keys, such that the last
value found will be used (except for the key \texttt{output}, see
below).  Table~\ref{tab:param_keys} presents the keys recognized by
the program.
If a key listed with a default value of ``\emph{none}'' is absent from
the parameter file, the program will print out a relevant error
message and the execution will be aborted.  The key \texttt{units} can
take the value \texttt{SI} if the \emph{Syst\`{e}me International} set of
units is desired (kg, m, s), with \texttt{AU} (the default)
corresponding to atomic units, where $\hbar = m_\rme = e = 1$, with
$m_\rme$ and $e$ the mass and the charge of the electron,
respectively.  Some equivalences between the two sets are given in
Tab.~\ref{tab:units}.  All parameters with units (mass, grid limits,
time step) must be consistent with the set of units chosen.
\newlength{\desc}
\setlength{\desc}{50mm}
\newlength{\default}
\setlength{\default}{30mm}
\begin{table}
  \caption{\label{tab:param_keys}Recognized parameters to be found in
    the parameter file.  Parameters with no default value must be
    present, with the exception of those indicated as \emph{none*}.}
  
  \begin{center}
    \begin{tabular}{|l|l|l|c|} \hline \multicolumn{1}{|c|}{Key} &
      \multicolumn{1}{c|}{Value type} &
      \multicolumn{1}{c|}{Description} & Default value \\ \hline \hline 
      \texttt{units} & \texttt{double}
      & System of units used, & \texttt{AU} \\
      & & SI or atomic units (AU) & \\ \hline
      \texttt{mass} & \texttt{double} & $m$, mass of the particle &
      \emph{none} \\ \hline 
      \texttt{nx} & \texttt{size\_t} & $n_x$, number
        of grid points & \emph{none} \\
      & & in $x$ & \\ \hline
      \texttt{ny} & \texttt{size\_t} & $n_y$, number
        of grid points & 1 \\
      & & in $y$ & \\ \hline
      \texttt{nz} & \texttt{size\_t} & $n_z$, number
        of grid points & 1 \\
      & & in $z$ & \\ \hline
      \texttt{x\_min} & \texttt{double} & Value of the first grid
      point & \emph{none} \\ 
      & & along $x$ & \\ \hline
      \texttt{x\_max} & \texttt{double} & Value of the last grid
      point & \emph{none} \\ 
      & & along $x$ & \\ \hline
      \texttt{y\_min} & \texttt{double} & Value of the first grid
      point & 0 \\
      & & along $y$ & (\emph{none} if $n_y > 1$) \\ \hline
      \texttt{y\_max} & \texttt{double} & Value of
        the last grid point & \texttt{y\_min} \\
      & & along $y$ &  (\emph{none} if $n_y > 1$) \\ \hline
      \texttt{z\_min} & \texttt{double} & Value of the first grid
      point & 0 \\
      & & along $z$ & (\emph{none} if $n_z > 1$) \\ \hline
      \texttt{z\_max} & \texttt{double} & Value of
        the last grid point & \texttt{z\_min} \\
      & & along $z$ &  (\emph{none} if $n_z > 1$) \\ \hline
      \texttt{dt} & \texttt{double} & Time step
      $\Delta t$ & \emph{none} \\ \hline
      \texttt{nt} & \texttt{unsigned int} & Number
      of time steps & \emph{none} \\ \hline
      \texttt{nprint} & \texttt{unsigned int}
      & Interval of the calculation & (see text) \\ 
      & & of the observables & \\ \hline
      \texttt{results\_file} & \texttt{char}
      & Output file name for
      & \texttt{results} \\
      & & observables & \\ \hline
      \texttt{wf\_output\_text} & \texttt{char}
      & File name for output of final & \emph{none*} \\
      & & wave function in text format & \\ \hline
      \texttt{wf\_output\_binary} & \texttt{char}
      & File name for output of final & \emph{none*} \\
      & & wave function in binary & \\
      & & format & \\
      \hline
    \end{tabular}
  \end{center}
\end{table}
\begin{table}
  \caption{\label{tab:units}Values of some atomic units~\cite{codata2010}.}
  
  \begin{center}
    \begin{tabular}{|c|c|c|} \hline
      Atomic unit & Symbol & SI value \\
      \hline \hline
      length & $a_0$ & $0.529\,177\,210\,92 \times 10^{-10}\ \mathrm{m}$ \\
      time & & $2.418\,884\,326\,502 \times 10^{-17}\ \mathrm{s}$ \\
      mass & $m_\rme$ & $9.109\,382\,91 \times 10^{-31}\ \mathrm{kg}$ \\
      energy & $E_{\mathrm{h}}$ & $4.359\,744\,34 \times 10^{-18}\
      \mathrm{J}$ \\ 
      \hline
    \end{tabular}
  \end{center}
\end{table}

In addition, the output of the program is controlled by a series of
flags, set in the same fashion as above, with the \emph{key}
\texttt{output} and \emph{value} equal to the desired flag. A list of
valid flags is given in Tab.~\ref{tab:param_output}.
%
%
\begin{table}
  \caption{\label{tab:param_output}Recognized output flags.}
  
  \begin{center}
    \begin{tabular}{|l|l|} \hline
      \multicolumn{1}{|c|}{Flag} & \multicolumn{1}{c|}{Description} \\
      \hline \hline 
      \texttt{norm} & Norm, $\sqrt{\braket{\psi|\psi}}$ \\
      \texttt{energy} & Energy, $E=\braket{\psi|\hat{H}|\psi}$ \\
      \texttt{x\_avg} & Average position in $x$,
      $\braket{x} = \braket{\psi|x|\psi}$ \\
      \texttt{y\_avg} & Average position in $y$,
      $\braket{y} = \braket{\psi|y|\psi}$ \\
      \texttt{z\_avg} & Average position in $z$,
      $\braket{z} = \braket{\psi|z|\psi}$ \\ 
      \texttt{sx} & Width in $x$, $\left\langle x^2 \right\rangle -
      \left\langle x \right\rangle^2$ \\  
      \texttt{sy} & Width in $y$, $\left\langle y^2 \right\rangle -
      \left\langle y \right\rangle^2$ \\  
      \texttt{sz} & Width in $z$, $\left\langle z^2 \right\rangle -
      \left\langle z \right\rangle^2$ \\  
      \texttt{autocorrelation} & Autocorrelation function, $\left|
        \Braket{\psi(0) | \psi(t)} \right|^2$ \\
      \texttt{user\_defined} & User-defined observables (see
      Sec.~\ref{sec:user})\\ 
      \hline
    \end{tabular}
  \end{center}
\end{table}
These values will be printed out in the file designated by the
\texttt{results\_file} key, for the initial wave function and every
\texttt{nprint} iteration of the time step $\Delta t$. The program
will abort with an error message if $\texttt{nprint} > \texttt{nt}$.
Note that if $\texttt{nt} \bmod \texttt{nprint} \neq 0$, the values
for the final wave function will not be calculated.  The key
\texttt{nprint} needs only be present if any of the output flags is
set.

\subsection{Memory usage}
\label{sec:memory}

Calculating the exact memory usage is a bit tricky, but as the main
use of memory is to store the wave function and some work arrays, we
can estimate a minimum amount of memory necessary according to the
grid size.  Considering that a double precision real takes up 8 bytes
of memory, the program requires at least
$$
40 \frac{\left( n_x n_y n_z \right)}{n_\mathrm{proc}} + 56 \left(
  \frac{n_x}{n_\mathrm{proc}} + n_y + n_z \right) 
$$
bytes \emph{per processor}, where $n_\mathrm{proc} \equiv \texttt{size}$ is
the number of processors used.  This value holds when the
autocorrelation function is not calculated; otherwise, the initial
wave function must be stored and the factor $40$ above changes to
$56$.  Obviously, this estimate does not include any memory allocated
within user-supplied routines.

\section{Sample results}
\label{sec:results}

\subsection{Laser excitation of vibration}
\label{sec:vib}

As a first example, let us consider a vibrating diatomic molecule,
with the Hamiltonian
\begin{equation}
  \hat{H} = -\frac{\hbar^2}{2 m} \frac{1}{r^2} \frac{\rmd}{\rmd r} r^2
    \frac{\rmd}{\rmd r} + \tilde{V}(r),
\end{equation}
for a wave function $\tilde{\psi}(r, \theta, \phi, t)$ in spherical
coordinates, with $m$ the reduced mass and $\tilde{V}(r)$ the
molecular potential~\cite{cohen-tannoudji92}.  We neglect here the
rotation of the molecule, and only look at the radial part of the wave
function, $\tilde{\psi}(r, t)$.  Setting $\psi \equiv r \tilde{\psi}$,
and substituting $x$ for $r$, we get the one-dimensional Schr\"{o}dinger
equation
\begin{equation}
  i \hbar \frac{\partial}{\partial t}  \psi(x, t) = \left[
    -\frac{\hbar^2}{2 m} \frac{\rmd^2}{\rmd x^2} + V(x,t) \right]
    \psi(x, t),
\end{equation}
which is the one-dimensional equivalent of eq.~(\ref{eq:tdse}) with
Hamiltonian eq.~(\ref{eq:hamiltonian}) and with the full potential
$V(x,t)$ taken as a sum of the molecular potential $\tilde{V}(x)$ and
the coupling of the molecule to a laser pulse, $V_{\mathrm{L}}(x,t)$.
We note that recovering an operator of the form $\rmd^2 / \rmd x^2$ is
a very special case obtained here for a diatomic molecule, and that in
general the kinetic energy operator for the internal motion of a
molecule can be quite different, such that this program may not be
used to study the internal dynamics of molecules in general.

For the molecular potential, we take a Morse
potential~\cite{morse29,atkins11},
\begin{equation}
  \tilde{V}(x) = D \left[ 1 - \rme^{-a (x - x_\rme)} \right]^2,
\end{equation}
and from the data of ref.~\cite{huber79}, we derive the parameters for
$^{12}$C$^{16}$O in the ground electronic state:
\begin{align*}
  m &= 12498.10 \\
  D &= 0.4076 \\
  a &= 1.230211\\
  x_\rme &= 2.1322214
\end{align*}
with $m = m_{\mathrm{C}} m_{\mathrm{O}}/(m_{\mathrm{C}} +
m_{\mathrm{O}}) $ the reduced mass, and all values expressed in atomic
units (see Tab.~\ref{tab:units}).

Using a classical model for the laser field and the dipole
approximation, the laser-molecule coupling is given by~\cite{milf}
\begin{equation}
V_{\mathrm{L}}(x,t) = \mu(x) \mathcal{E}(t),
\end{equation}
where $\mu(x)$ is the dipole moment of the molecule and
$\mathcal{E}(t)$ the electric field of the laser.  We approximate the
internuclear-separation-dependent permanent dipole moment of the
molecule as the linear function
\begin{equation}
\mu(x) = \mu_0 + \mu' \left( x-x_\rme \right),
\end{equation}
with the values (in atomic units) $\mu = -0.1466$ and $\mu' =
-0.948$~\cite{hcn:gready78}.  For the laser pulse, we take
\begin{equation}
  \mathcal{E}(t) = \mathcal{E}_0 f(t) \cos ( \omega t )
\end{equation}
with $\mathcal{E}_0$ and $\omega$ the amplitude and frequency of the
field, respectively, and the envelope function
\begin{equation}
  f(t) = 
  \begin{cases}
    \sin^2 \left( \pi \frac{t}{t_{\mathrm{f}} - t_{\mathrm{i}}}
    \right) & \text{if
      $t_{\mathrm{i}} \le t \le t_{\mathrm{f}}$} \\
    0 & \text{otherwise}
    \end{cases}
\end{equation}
In this sample simulation, we take the following values (in atomic
units):
\begin{align*}
  \mathcal{E}_0  &= 1.69 \times 10^{-3} \\
  \omega &= 9.8864 \times 10^{-3} \\
  t_{\mathrm{i}} &= 0 \\
  t_{\mathrm{f}} &= 41341.37
\end{align*}
This corresponds to a 1~ps pulse at an irradiance of $10^{11}\
\mathrm{W/cm}^2$, resonant with the $v=0 \rightarrow v=1$ transition.

Using a \textsc{dvr} method~\cite{dvr:colbert92}, we precomputed the
first five vibrational eigenstates $\phi_v$ of the Morse potential for
$^{12}$C$^{16}$O on a grid of 4000 points, from $x = 1.5 \times
10^{-3}\ \mathrm{a.u.}$ to $6\ \mathrm{a.u.}$.  The data, stored in
file \texttt{CO\_vib.txt}, are read when the wave function is
initialized in the function \texttt{initialize\_wf}, and the initial
wave function is set to $\psi(x,t=0) = \phi_0(x)$.  The function
\texttt{user\_observe} is programmed to calculate the projection of
the wave function on the first five eigenstates, \ie,
\begin{equation}
  \mathcal{P}_v(t) \equiv \left| \braket{ \phi_v | \psi(t)} \right|^2.
\end{equation}

\begin{figure}
  \centerline{\includegraphics{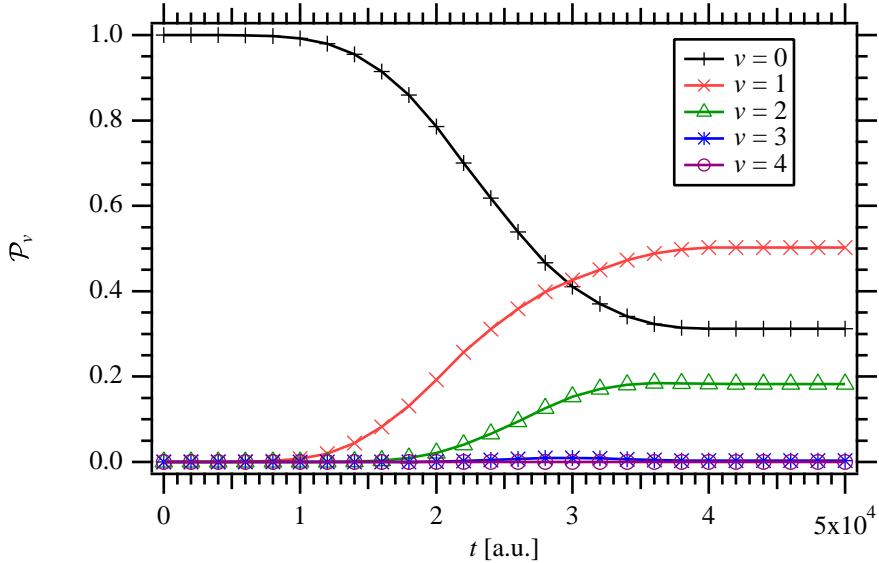}}
  \caption{\label{fig:co_proj}Projection of the time-dependent
    vibrational wave function of the CO molecule, interacting with a
    resonant laser pulse, on the first five vibrational eigenstates.}
\end{figure}

Using the same grid as the one described above for the calculation of
the vibrational states, we run the simulation for $500\,000$ time
steps of length $\Delta t = 0.1\ \mathrm{a.u.}$, and calculate the
projection of the wave function on the vibrational eigenstates every
$20\,000$ time steps.  the result is shown in fig.~\ref{fig:co_proj}.

\subsection{Atomic ion in a Paul trap}

Let us now consider the three-dimensional problem of the motion of a
charged atomic ion in a Paul
trap~\cite{Paul_RMP_1990,Major_book_2005,Werth_book_2009}.  These create a
time-dependent quadrupolar field allowing, under the right conditions,
the confinement of an ion.

The electric potential inside a Paul trap is of the
form~\cite{Major_book_2005,Werth_book_2009}
\begin{equation}
  \Phi(\mathrm{x},t) = \frac{U_0 + V_0 \cos \Omega t}{2 d^2} \left(
    r^2 - 2 z^2 \right),
\end{equation}
where $U_0$ is a static electric potential, $V_0$ the amplitude of an
ac potential of frequency $\Omega$, and $r^2 \equiv x^2 + y^2$.  The
scale factor $d$ is obtained from $d^2 = r_0^2 + 2 z_0^2$, with $r_0$
the radial distance from the center of the trap to the ring electrode
and $z_0$ the axial distance to an end cap (see
refs.~\cite{Major_book_2005,Werth_book_2009} for more details).
Considering an atomic ion of charge $Ze$, where $e$ is the elementary
charge~\cite{codata2010}, we get the potential energy
\begin{equation}
  V(\mathbf{x},t) = Z e \Phi(\mathrm{x},t).
\end{equation}

\begin{figure}
  \begin{center}
    \includegraphics{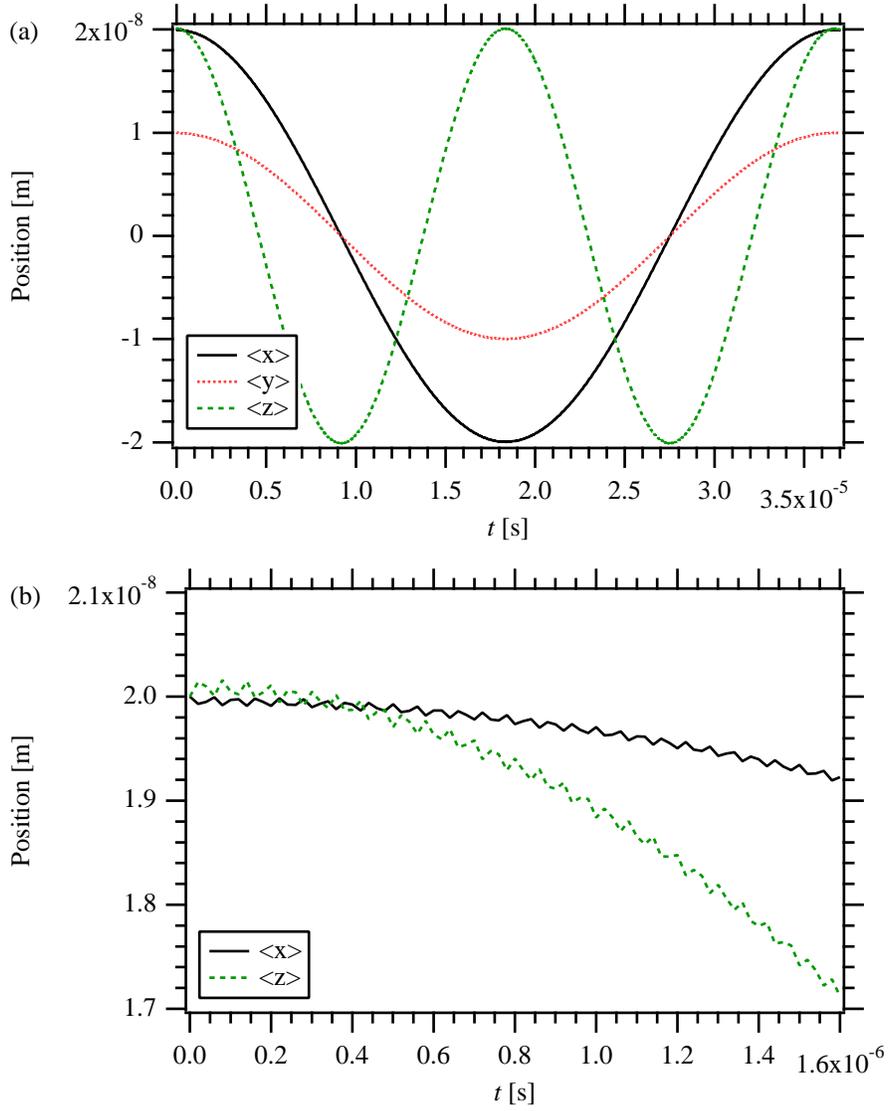}    
  \end{center}
  \caption{\label{fig:traj}(a) Sample trajectory of the wave packet of
    a Ba$^+$ ion in a Paul trap.  The simulation is carried in three
    dimensions, and the expectation value of the position is plotted
    individually for each Cartesian coordinate.  (b) Enlargement of
    panel (a), evidencing the micromotion of the ion at the frequency
    of the trapping potential.}
\end{figure}

For the simulation, we consider conditions similar to those of
refs.~\cite{neuhauser_prl_1978,neuhauser_ap_1978} and take a
$^{138}$Ba$^{+}$ ion, $m= 137.905232\ \mathrm{u} = 2.28997005 \times
10^{-25}\ \mathrm{kg}$~\cite{nist:atomic}, in a trap with
characteristics:
\begin{align*}
  U_0 &= 0\ \mathrm{V} \\
  V_0 &= 200\ \mathrm{V} \\
  \Omega &= 2\pi \times 18\ \mathrm{MHz}\\
  r_0 &= 1.6 \times 10^{-3}\ \mathrm{m} \\
  z_0 &= r_0 / \sqrt{2}
\end{align*}
The initial state is taken as a Gaussian wave packet,
\begin{equation}
  \psi_\rmi(x,y,z) =  \left(\frac{2}{\pi}\right)^{3/4} \prod_{\xi =
    x,y,z} \frac{1}{\sqrt{\sigma_\xi}} \exp \left[ \frac{\rmi}{\hbar}
    p_{\xi0} \left( \xi-\xi_0 \right)  \right] \exp \left[-
    \frac{\left( \xi-\xi_0 
      \right)^2}{\sigma_\xi^2} \right],
\end{equation}
and we set
\begin{align*}
  x_0 &= z_0 = 2 \times 10^{-8}\ \mathrm{m} \\
  y_0 &= 1 \times 10^{-8}\ \mathrm{m} \\
  p_{x0} &= 1 \times 10^{-27}\ \mathrm{kg\,m\,s}^{-1} \\
  p_{y0} &= p_{z0} = 0 \\
  \sigma_x &= \sigma_y = 7.342 \times 10^{-8}\ \mathrm{m} \\
  \sigma_z &= 5.192  \times 10^{-8}\ \mathrm{m}
\end{align*}
Using $\texttt{nx} = \texttt{ny} = \texttt{nz} = 512$ grid points,
with the grid defined from $-1 \times 10^{-6}\ \mathrm{m}$ to $1
\times 10^{-6}\ \mathrm{m}$ along each Cartesian coordinate, we run
the simulation for $\texttt{nt} = 18\,500$ time steps of length
$\Delta t = 2 \times 10^{-9}\ \mathrm{s}$, measuring the wave function
every 10 time steps.  The resulting trajectory of the ion is shown in
fig.~\ref{fig:traj}.


\section*{Acknowledgements}

This research was conducted using the resources of the High
Performance Computing Center North (HPC2N).  Funding from Ume{\aa}
University is gratefully acknowledged.






\end{document}